\documentstyle[fleqn,epsf]{article}
\title{\bf Diffusion-annihilation dynamics 
 in one spatial dimension}
\author{J. E. Santos, G. M. Sch\"{u}tz and 
 R. B. Stinchcombe\\
 Department of Physics, University of Oxford\\
 Theoretical Physics, 1 Keble Road,
 Oxford OX1 3NP, UK\\Email address: jesantos@thphys.ox.ac.uk}
\date{}
\newcommand{\beq}{\begin{equation}}
\newcommand{\eeq}{\end{equation}}
\newcommand{\ket}[1]{\mbox{$ \mid #1\, \rangle$}}
\newcommand{\bra}[1]{\mbox{$ \langle\, #1\mid$}}
\newcommand{\noc}[1]{\mbox{$\hat{n}_{#1}$}}
\newcommand{\spup}[1]{\mbox{$\hat{s}^{+}_{#1}$}}
\newcommand{\spdo}[1]{\mbox{$\hat{s}^{-}_{#1}$}}
\newcommand{\crea}[2]{\mbox{$\hat{#1}^{\dagger}_{#2}$}}
\newcommand{\etal}[1]{\mbox{\normalsize 
$\eta_{\mbox{\scriptsize $#1$}}$}}
\newcommand{\anni}[2]{\mbox{$\hat{#1}_{#2}$}}

\begin{document}
\maketitle
\begin{abstract}
 We discuss a reaction-diffusion model in  
 one dimension subjected to an 
 external driving force. Each lattice site may 
 be occupied by at most one 
 particle. The particles hop with rates $(1\pm\eta)/2$ 
 to the right or left 
 nearest neighbour site if it is vacant, and 
 annihilate with rate one if it is 
 occupied. We compute the long time behaviour 
 of the space dependent average 
 density in states where the initial density 
 profiles are step functions. 
 We also compute the exact time dependence of 
 the particle density for 
 uncorrelated random initial conditions. The 
 representation of the uncorrelated 
 random initial state and also of the step 
 function profile in terms of free 
 fermions allows for the calculation of 
 time-dependent higher order correlation 
 functions. We outline the procedure 
 using a field theoretic approach. 
 Finally, we show how this gives rise to 
 predictions on experiments in TMMC exciton dynamics.
\end{abstract}
\pagebreak
\section{Introduction}
\label{sec1}
 In recent years there has been much 
 interest in one-dimensional 
 reaction-diffusion systems. Despite 
 their simplicity, they exhibit
 a very rich dynamical behaviour and 
 have been used as models for 
 experimental systems \cite{r1}. Among 
 them, the study of diffusion 
 of hard-core particles (e.g. with exclusion) 
 plays a central role 
 \cite{r2,r2a,r2b}. Such systems are usually 
 described in terms of 
 classical (mean field) equations. However, 
 in low dimensional 
 systems, where the diffusion of the 
 reactants is not sufficiently 
 rapid compared to the reaction rate, 
 large fluctuations persist 
 and lead to a breakdown of the mean 
 field approximation. Thus 
 more rigorous methods are needed, e.g. 
 a description of the 
 model in terms of a master equation.

 In this paper we will consider a model 
 which allows for the 
 full solution of the master equation. 
 It is defined on a ring 
 of \(L\) sites  where each lattice site 
 may be occupied by 
 at most one particle. This exclusion 
 principle models a 
 hard-core repulsion of the particles. 
 These particles 
 (which will be denoted by \(A\)) hop 
 with rates 
 \( (1 \pm \eta)/2\) to the right or 
 left nearest 
 neighbouring sites respectively, if the 
 site is vacant 
 (denoted by 
 \(\emptyset\)), and annihilate with rate \(\lambda\) 
 if it is occupied:
\beq
A \: \emptyset \: \longrightarrow \: \emptyset \: 
A \;\;\;\;\; (1 + \eta)/2
\label{1}
\eeq
\beq
\emptyset \: A \: \longrightarrow \: A \: \emptyset 
\;\;\;\;\; (1 - \eta)/2
\label{2}
\eeq
\beq
A \: A \: \longrightarrow \: \emptyset \: \emptyset  
\;\;\;\;\; \lambda .
\label{3}
\eeq

 We will study \(\lambda=1\). This  case  is 
 the one which 
 is fully solvable, and besides the results 
 obtained are 
 of experimental relevance for the study of 
 exciton dynamics 
 in TMMC polymers (see below). This model, 
 which is closely 
 related to zero temperature Glauber 
 dynamics \cite{r3}, 
 has already been 
 studied by many authors, see e.g. \cite{r7} - \cite{r11}.

 It turns out to be convenient to write 
 the master equation 
 of the process in 
 terms of a quantum spin Hamiltonian. 
 Since the particle 
 model involves two 
 states per site, it maps into a spin 1/2 system, 
 the evolution operator of 
 which is the quantum spin Hamiltonian. 
 The Jordan-Wigner transformation allows 
 this Hamiltonian to be written in terms of fermion 
 operators and reduces for 
 \(\lambda=1\) to a free fermion system. 
 This somewhat surprising simplification 
 allows for the exact calculation of many 
 dynamical properties of the system.
 In particular it has been known for a long 
 time that the density for an 
 initially completely filled state decays 
 as \(t^{-1/2}\) \cite{r7} corresponding 
 to a diffusive ``critical'' behaviour.

 We extend these results and calculate 
 explicitly the time evolution of the
 density for uncorrelated random initial 
 states with any given density 
 \(0\leq\rho\leq 1\). We also show how to 
 calculate higher order correlation 
 functions. Moreover, we study the time 
 evolution of the space dependent average 
 density for non-translationally invariant 
 initial states, namely the step 
 function and  half step function in the 
 density profile. With this we obtain 
 the development in time of the reaction 
 edge of a system which is initially 
 completely empty to the left of the origin and 
 full (half-full) to the right of 
 it. Also, we have found a convenient way 
 to calculate time-dependent higher 
 order correlation functions for the 
 step function initial state.

 The interest of studying this problem is not 
 merely theoretical. This model 
 can be mapped by a similarity transformation 
 into the coagulation problem where 
 we have the reaction  (\(A \: A \: \longrightarrow 
 \: A \: \emptyset \: , 
 \: \emptyset\: A\)) rather than the pair 
 annihilation \(A \: A \: \longrightarrow \: 
 \emptyset \: \emptyset \:\) 
 \cite{Krebs}. It has been suggested \cite{r10,r11}
 to use this lattice model 
 for the description of exciton kinetics 
 in the carrier substance 
 \(N(CH_{3})_{4}MnCl_{3}\) (TMMC) \cite{r1}: 
 Excitons of the \(Mn^{2+}\) ion are 
 generated after pulsed laser excitations of 
 the sample and diffuse along the 
 widely separated \(MnCl_{3}\) chains. 
 Experimental data show that a single 
 exciton has a decay time of about \(0.7 ms\). 
 The on-chain hopping rate is 
 \(10^{11}-10^{12} s^{-1}\). If two excitons 
 arrive on the same \(Mn^{2+}\) ion, 
 they undergo a coagulation reaction 
 (\(A + A \: \longrightarrow \: A \)), under 
 the emission of radiation. The reaction 
 time for this process is assumed to
 be \(\approx 10^{-13} s\) which is much 
 faster than the diffusion time. 
 Measurements of the time resolved 
 luminescence in TMMC, in  a  
 time scale of a picosecond to a millisecond 
 allow the observation of the decay 
 of the exciton population. Two characteristics 
 of this system are especially 
 relevant to the lattice coagulation model 
 we are using to describe it. 
 Firstly, given that the reaction process is 
 much faster than the diffusion 
 process, the coagulation rate for the lattice 
 model is determined to be equal 
 to the hopping rate: If two
 particles (i.e. excitons) are on neighbouring
 lattice sites (i.e. \(Mn^{2+}\)
 ions), either of them will hop onto the site
 occupied by the other with the 
 (unit) diffusion rate. Almost instantaneously
 one of them will annihilate and 
 as a result one particle will be left,
 with equal probability on either site.
 Secondly, the single exciton decay is
 very slow compared with the other 
 processes and we can, as a first approximation
 to the dynamics, neglect it.
 The model then exactly maps into the
 annihilation problem with 
 \(\lambda=1\). Hence, we can obtain expressions
 for physical quantities that 
 can be compared with experiments.

 The structure of this paper is as follows:
 In section 2 we introduce the 
 description of this system in terms of
 the master equation and analyse its 
 mapping to the quantum spin problem. In section
 3, we sketch how this problem
 can be converted into a problem of spinless fermions,
 by means of the Jordan-Wigner
 transformation, and how to calculate the
 relevant quantities in this fermion
 language. We proceed to the calculation
 of the density for the uncorrelated states,
 in section 4. We postpone to Appendix A the
 techniques which allow the calculation 
 of higher order correlation functions. In
  section 5 we present the calculation of 
 the average space dependent density for
 the step  and half step function states, 
 and again leave the 
 details related to the calculation of
 higher order correlation functions to an appendix,
 in this case Appendix B. We repeat the former
 calculation for a half step function state.
 In section 6, we briefly review the concept
 of similarity transformation, and use it 
 to calculate quantities which can be compared
 with experiments. Finally, in section 7, we present
 our conclusions and
 some open questions.

\section{The master equation and the mapping 
 to the spin system}
\label{sec2}

 Our system is defined on a 
 lattice of \(L \) sites with periodic boundary
 conditions.
 Each site has assigned a particle occupation
 number \(n_{j}=0,1\;
 \;(1\leq j \leq L) \). The set \(\underline{n}\:=\;\{n_{1},
 \ldots,n_{L}\}\) 
 characterizes a 
 given configuration of particles. The process is
 completely defined by the
 probability distribution \(P(\underline{n},t)\)
 which gives the probability
 for the system to be in the configuration
 \(\underline{n}\) at time \(t\). 
 \(P(\underline{n},t)\) satisfies a master
 equation which is conveniently
 expressed in terms of a quantum Hamiltonian
 language \cite{r8}.

 The idea is to assign to each of the
 \(2^{L}\) configurations of this system a 
 normalized basis vector $\ket{\underline{n}}$
 in a Hilbert space. The 
 probability distribution is then
 mapped to a state vector
\beq
\ket{P(t)}\;=\;\sum_{\underline{n}\in X}\; 
P(\underline{n},t)\;\ket{\underline{n}}.
\label{4}
\eeq

 Using this notation it is then 
 possible to write the master equation in a
 ``Schr\"{o}dinger form''
\beq
\frac{\partial}{\partial t}\,\ket{P(t)}\;=\;-\;H\;\ket{P(t)}
\label{5}
\eeq
 where \(H\) is a linear, in general 
 non-hermitian operator, containing all
 the information on the stochastic process.
 Equation (\ref{5}) has the formal solution
\beq
\ket{P(t)}\;=\;\exp\,(-\,H\,t)\;\ket{P(0)}.
\label{6}
\eeq

 Now one can express the time dependent 
 averages of any function
 \(A(n_{1},\ldots,n_{L})\) of the occupation 
 numbers as
\beq
\langle A (n_{1},\ldots,n_{L}) \rangle\;=\;
\bra{s}\,A(\noc{1},\ldots,\noc{L})\,\ket{P(t)}
\label{7}
\eeq
 where the \noc{j}'s are the projection operators 
 on states with a particle
 on site \(j\)
 of the chain and \bra{s} is defined as
\beq 
\bra{s}\;=\;\sum_{\underline{n}\in X}\;\bra{\underline{n}}.
\label{8}
\eeq
 Using (\ref{5}) one sees that
\beq
\bra{s}\;H\;=\;0.
\label{9}
\eeq
 This expresses the conservation of the 
 total probability.

 Since we are working with a two state model 
 the projection operators
 \noc{j} can be represented in terms of 
 Pauli matrices
\beq
\noc{j}\;=\;\frac{1}{2}\,(1-\hat{\sigma}^{z}_{j})\;=\;
\left(\begin{array}{cc}
 0&0\\
 0&1
\end{array}\right)_{j}.
\label{10}
\eeq
 Also, for convenience, we introduce 
 the operators
\(\hat{s}^{\pm}_{j}\;=\;(\,\hat{\sigma}^{x}_{j}\;
\pm\;\mbox{i}\,\hat{\sigma}^{y}_{j}\,)/2\).
 In the convention used just above
\beq
\spdo{j}\;=\;
\left(\begin{array}{cc}
 0&0\\
 1&0
\end{array}\right)_{j}
\label{11}
\eeq
 creates a particle at site \(j\), and
\beq
\spup{j}\;=\;
\left(\begin{array}{cc}
 0&1\\
 0&0
\end{array}\right)_{j}
\label{12}
\eeq
 annihilates a particle at site \(j\).

 With these operators the quantum Hamiltonian
 which describes the stochastic dynamics
 of the pair annihilation process is given by
\begin{eqnarray}
 H&=&\sum_{j=1}^{L}\;\{\,\frac{1}{2}\,(1+\eta)
\,[\,\noc{j}\,
 (1-\noc{j+1})\;-\;\spup{j}\,\spdo{j+1}\,]
\nonumber \\
       & &\mbox{} +\frac{1}{2}\,(1-\eta)\,
[\,(1-\noc{j})\,\noc{j+1}
\;-\;\spdo{j}\,\spup{j+1}\,]
\nonumber \\
       & &\mbox{} 
+\lambda(\,\noc{j}\,\noc{j+1}\;-\;\spup{j}
\,\spup{j+1}\,)\,\}.
\label{13}
\end{eqnarray}

 We conclude this section noting that we 
 can change from the Schr\"{o}dinger 
 representation of the operators to the
 Heisenberg one in a similar way as in 
 quantum mechanics. The time dependent
 operators are defined by
\beq
 A(t)\;=\;\exp\,(\,H\,t)\;A\;\exp\,(-\,H\,t)
\label{14}
\eeq
 and obey the equation of motion
\beq
\frac{d\,A(t)}{d\,t}\;=\;[\,H\, , \, A(t)\,].
\label{15}
\eeq

\section{The fermion representation of the operators}
\label{sec3}

 We introduce a fermion representation of the 
 Hamiltonian by means of the Jordan-Wigner
 transformation \cite{r9}. Defining
\beq
\hat{Q}_{j}\;=\;\prod_{i=1}^{j}\,\hat{\sigma}_{i}^{z}
\label{16}
\eeq
\beq
\crea{a}{j}\;=\;\spdo{j}\,\hat{Q}_{j-1}
\label{17}
\eeq
\beq
\anni{a}{j}\;=\;\hat{Q}_{j-1}\,\spup{j}
\label{18}
\eeq
 it is a simple exercise to show that the set 
 of \crea{a}{j}'s,
\anni{a}{l}'s obey the fermion anticommutation 
 relations
\begin{eqnarray}
\{\,\anni{a}{j}\, , \,\anni{a}{l}\,\}&=&\{\,\crea{a}{j}
\, , \,\crea{a}{l}\,\}\;=\;0 \nonumber\\
\{\,\anni{a}{j}\, , \,\crea{a}{l}\,\}&=&
\delta_{j,l}.
\label{19}
\end{eqnarray}

 Note that because of the periodic boundary conditions
 for the spin operators one
 has \(\crea{a}{L+1}\:=\:\crea{a}{1}\,\hat{Q}_{L}\)
 and  \(\anni{a}{L+1}\:
 =\:\hat{Q}_{L}\,\anni{a}{1}\).
 \(\hat{Q}_{L}\) may be written \(\hat{Q}_{L}
 \:=\:(-1)^{\hat{N}}\),
 where \(\hat{N}\:=\:\sum_{j}\,\noc{j}\)
 is the number operator. Since the action 
 of \(H\) changes the particle
 number only by units of two, \(\hat{Q}_{L}\) 
 is a constant of motion and splits 
 the Hilbert space into a sector with 
 an even number of particles 
 (\(\hat{Q}_{L}\:=\:1\)) and into a sector 
 with an odd number of
 particles (\(\hat{Q}_{L}\:=\:-1\)). The 
 application of \anni{a}{L+1} into states 
 belonging to one of this sectors gives 
 \(\anni{a}{L+1}\,\ket{X}\:=\:-\,\anni{a}{1}\,\ket{X}\) 
 if \(X\:\in\) even sector and
 \(\anni{a}{L+1}\,\ket{X}\:=\:\anni{a}{1}\,\ket{X}\)
 if \(X\:\in\) odd sector. 
 This implies that we have antiperiodic (periodic)
 boundary conditions for the 
 even (odd) sector. Applying the equations
 (\ref{16}) - (\ref{18}) to the 
 Hamiltonian (\ref{13}), one obtains \cite{r5}
\begin{eqnarray}
 H&=&-\;\frac{1}{2}\:\sum_{j=1}^{L}\,
 (\,\crea{a}{j+1}\,\anni{a}{j}\; 
 +\;\crea{a}{j}\,\anni{a}{j+1}
 \;+\;2\,\anni{a}{j+1}\,\anni{a}{j}\;-\;2\,\noc{j}\,)
 \;-\;(\,\lambda\,-\,1\,)\,
 \sum_{j=1}^{L}\,(\,\anni{a}{j+1}\,\anni{a}{j}\;
 -\;\noc{j}\,\noc{j+1}\,)
\nonumber \\
       & & \mbox{} -\;\frac{\eta}{2}\,\sum_{j=1}^{L}
\,(\,\crea{a}{j+1}\,\anni{a}{j}\;
 -\;\crea{a}{j}\,\anni{a}{j+1}\,)
\label{20}
\end{eqnarray}
(\(\noc{j}\:=\:\crea{a}{j}\,\anni{a}{j}\)).
 Notice that when we take
 \(\lambda\:=\:1\), the
 two body term vanishes, making this
 Hamiltonian a free fermion one. 

 We now define the Fourier transforms
\beq
\anni{b}{p}\;=\;\frac{e^{-i\frac{\pi}{4}}}{\sqrt{L}}\,
\sum_{j=1}^{L}\,e^{\frac{2\pi i}{L}pj}\,\anni{a}{j}
\label{21}
\eeq
\beq
\crea{b}{p}\;=\;\frac{e^{i\frac{\pi}{4}}}{\sqrt{L}}\,
\sum_{j=1}^{L}\,e^{-\frac{2\pi i}{L}pj}\,\crea{a}{j}
\label{22}
\eeq
 where \(p\) runs over all integers 
 \(p\:=\:-\frac{L}{2},\ldots,0,\ldots,\frac{L}{2}-1\)
 in the sector with odd number of particles
 and over the half-odd integers \(p\:=\:-\frac{L}{2}-
\frac{1}{2},\ldots,\frac{1}{2},\frac{3}{2},\ldots,
\frac{L}{2}-\frac{1}{2}\)
 in the even sector (we take \(L\) even).
 Also, the transformation (\ref{21}) - (\ref{22})
 is canonical and one has \(\{\,\anni{b}{p}\, , \,
 \anni{b}{q}\,\}\:=\:\{\,\crea{b}{p}\, , \,\crea{b}{q}
 \,\}\:=\:0 \) and
 \(\{\,\anni{b}{p}\, , \,\crea{b}{q}\,\}\:=\:\delta_{p,q}\).
 In terms of the
 \anni{b}{p}'s, \crea{b}{q}'s, the Hamiltonian
 (\ref{20}) reads \cite{r5}
\begin{equation}
 H\;=\;\sum_{p}\, \left[\,\left(\,1\:-\:\cos\,
\left(\,\frac{2\pi p}{L}\,\right)\,\right)\,
\crea{b}{p}\,\anni{b}{p}
\;+\;\sin\,\left(\,\frac{2\pi p}{L}\,\right)
\,\anni{b}{-p}\,\anni{b}{p}\,\right]
\;-\;\mbox{i}\,\eta\,\sum_{p}\,\sin\,
\left(\,\frac{2\pi p}{L}\,\right)\,
\crea{b}{p}\,\anni{b}{p}.
\label{23}
\end{equation}

 Since \(H\) is quadratic in the \anni{b}{p}'s, 
 \crea{b}{q}'s, the equations of motion
 (\ref{15}) for these operators are linear
 and become easily solvable \cite{r5}. Their solution is
\beq
\anni{b}{p}(t)\;=\;e^{-\epsilon_{p}t}\,\anni{b}{p}
\label{24}
\eeq
\beq
\crea{b}{p}(t)\;=\;e^{\epsilon_{p}t}\,\left[\,\crea{b}{p}
\;+\;\cot\,\left(\,\frac{\pi p}{L}\,\right)\,(\,1\;
 -\;e^{-(\epsilon_{p}
 +\epsilon_{-p})t}\,)\,\anni{b}{-p}\,\right]
\label{25}
\eeq
 with
\beq
\epsilon_{p}\;=\;1\;-\;\cos\,\left(\,
\frac{2\pi p}{L}\,\right)\;-\;\mbox{i}\,\eta\,
\sin\,\left(\,\frac{2\pi p}{L}\,\right).
\label{26}
\eeq

 Having obtained the dynamical solution of
 the problem, we now turn to the 
 description of the initial conditions in
 this fermion language. With the 
 convention we adopted in section 2,
 the state with all spins up is
 the empty state, i.e. the vacuum. A state
 with particles in sites
\(j_{1}<j_{2}<\ldots<j_{N}\;(N\leq L)\) 
 is represented by
\beq
\ket{j_{1},j_{2},\ldots,j_{N}}\;=\;\spdo{j_{1}}\,
\spdo{j_{2}}\,\ldots\,\spdo{j_{N}}\,\ket{0}.
\label{27}
\eeq

 As \(j_{1}<j_{2}<\ldots<j_{N}\), \(\hat{Q}_{j_{1}-1}\)
 can be commuted with all the
 remaining \spdo{j} operators and applied to the
 vacuum, giving one. The same holds
 for the remaining \(\hat{Q}\) operators, giving
\beq
\ket{j_{1},j_{2},\ldots,j_{N}}\;=\;\crea{a}{j_{1}}\,
\crea{a}{j_{2}}\,\ldots\,\crea{a}{j_{N}}\,\ket{0}
\label{28}
\eeq
 i.e. the states are just Slater determinants
 in the position representation.
 Also, from the condition (\ref{9})
 \(\bra{s}\,H\;=\;0\), one obtains
\begin{eqnarray}
\bra{s}&=&\bra{0}\,\prod_{p}\,(\,1\;+\;\cot\,
\left(\,\frac{\pi p}{L}\,\right)\,\anni{b}{p}\,\anni{b}{-p}\,)\;
 +\;\bra{0}\,\anni{b}{0}\,\prod_{p'}\,(\,1\;+\;\cot\,
\left(\,\frac{\pi p'}{L}\,\right)\,\anni{b}{p'}\,\anni{b}{-p'}\,) 
\nonumber\\
 &=&\bra{0}\,\exp\,(\sum_{p}\,\cot\,\left(\,
\frac{\pi p}{L}\,\right)\,\anni{b}{p}\,\anni{b}{-p}\,)\;
 +\;\bra{0}\,\anni{b}{0}\,\exp\,(\sum_{p'}\,
\cot\,\left(\,\frac{\pi p'}{L}\,\right)\,\anni{b}{p'}\,
\anni{b}{-p'}\,)
\nonumber\\
 &=&\;\bra{s}^{even}\;+\;\bra{s}^{odd}.
\label{29}
\end{eqnarray}
 The product and sum  over \(p\), \(p'\) run over 
 \(p\:=\:\frac{1}{2},\frac{3}{2},\ldots,
\frac{L}{2}-\frac{1}{2}\) 
 (even sector) and over \(p'\:=\:0,\ldots,
\frac{L}{2}-1\) (odd sector),
 respectively and \(\bra{s}^{even(odd)}\) obey
\(\bra{s}^{even(odd)}\:H\:=\:0\) separately.

 For simplicity, we only consider states
 belonging to the even sector of the 
 Hilbert space. Since the calculation of
 all correlation functions implies
 the action of an even number of fermion
 operators to such a state, and then
 the contraction with \bra{s}, one can
 replace \bra{s} just by 
\(\bra{s}^{even}\) because the contraction
 with \(\bra{s}^{odd}\) gives 0.
 One property of \(\bra{s}^{even}\) which is useful
 to note for studying correlation
 functions is \cite{r5}
\beq
\bra{s}^{even}\,(\,\crea{b}{p}\;+\;\cot\,\left(\,
\frac{\pi p}{L}\,\right)\,\anni{b}{-p}\,)\;=\;0.
\label{30}
\eeq
 The expression (\ref{30}) allows the reduction
 of the calculation of any 
 correlation function to the calculation of
 linear combinations of quantities
\(\bra{s}\,\anni{b}{p_{1}}\ldots\anni{b}{p_{2N}}\,
\ket{\Psi}\) where the 
 coefficients are now functions of time (\ket{\Psi}
 is the initial state, and 
 belongs to the even sector). There is
 an advantage in
 working with a quantity like $\bra{s}\,
\anni{b}{p_{1}}\ldots\,\anni{b}{p_{2N}}\,
\ket{\Psi}$, namely its decomposition in
 terms of contractions of pairs, valid 
 for some sets of initial states
 (see appendices A and B).
 This is particularly useful for
 the calculation of higher order
 correlation functions in those states.
 So, for the calculation of multi-time
 correlations functions like
 \(\bra{s}\,\noc{j_{1}}(t_{1})\ldots
 \noc{j_{N}}(t_{N})\,\ket{\Psi}\), one may 
 use \(\noc{j}(t)\:=\:\crea{a}{j}(t)\,
 \anni{a}{j}(t)\) and express
 \(\crea{a}{j}(t)\) and \(\anni{a}{j}(t)\) 
 in terms of
 \(\crea{b}{p}(t)\) and \(\anni{b}{p}(t)\) 
 using (\ref{21}) and (\ref{22}).
 Then one expresses these operators in 
 terms of \crea{b}{p} and \anni{b}{p}
 with the help of (\ref{24}) and (\ref{25}). 
 Finally, one anticommutes all
 the \crea{b}{p}'s to the right and uses (\ref{30}).
 Hence, one obtains a sum 
 of strings which just involve the
 \anni{b}{p} operators.

 The strategy for the calculation of higher
 order correlators is discussed
 in the appendices. In what follows we study
 only the time evolution of the
 average density at a given site $j$ which,
 using the results presented above, 
 is given by
\beq
\langle\noc{j}(t)\rangle\;=\;\frac{1}{L}\,\sum_{p,p'}\,
 e^{\frac{2\pi i}{L}j(p-p')-(\epsilon_{-p}
 +\epsilon_{p'})t}\,\cot\,
\left(\,\frac{\pi p}{L}\,\right)\,
\bra{s}\,\anni{b}{p'}\,\anni{b}{-p}\,\ket{\Psi}
\label{31}
\eeq
 or, with the help of (\ref{30}),
\beq
\langle\noc{j}(t)\rangle\;=\;\frac{1}{L}\,\sum_{p,p'}\,
 e^{\frac{2\pi i}{L}j(p-p')-(\epsilon_{-p}
 +\epsilon_{p'})t}\,
\bra{s}\,\crea{b}{p}\,\anni{b}{p'}\,\ket{\Psi}.
\label{32}
\eeq

\section{The average density for states 
 with random initial conditions}
\label{sec4}

 Here we study the time evolution of the 
 average density for uncorrelated random 
 initial conditions, i.e. an initial state 
 \ket{\rho} where on each lattice site 
 the probability of finding a particle is 
 \(\rho\). Such a state is a 
 superposition of states with both even 
 and odd number of particles. So, one 
 projects it over the even and odd sectors, obtaining 
\beq
\ket{\rho}^{even(odd)}\:=\:\frac{2}{1\,\pm \,(1-2\rho)^{L}}\:
\frac{1\,\pm\,\hat{Q}_{L}}{2}\,\ket{\rho}
\label{33}
\eeq
 where \(\frac{1\,\pm\,\hat{Q}_{L}}{2}\) are
 the projectors over the even and 
 odd subspace and \(\frac{2}{1\,\pm \,(1-2\rho)^{L}}\)
 is a normalization factor.
 In the fermion representation the state
 \(\ket{\rho}^{even}\) is given by
\beq
\ket{\rho}^{even}\:=\:\frac{2\,(1-\rho)^{L}}{1\,+
\,(1-2\rho)^{L}}\,\prod_{p>0}\,
(1\:+\:\mu^{2}\,\cot\,\left(\,\frac{\pi p}{L}
\,\right)\,\crea{b}{-p}\,\crea{b}{p}\,)\,
\ket{0}
\label{34}
\eeq
 where \(\mu\:=\:\rho/(1-\rho)\) \cite{r5}.

 The form (\ref{34}) is particularly convenient
 for the calculation of average 
 values like \(\bra{s}\,\anni{b}{p'}\,
\anni{b}{p}\,\ket{\rho}^{even}\) because 
 of its simple form in momentum space.
 If we use representation (\ref{29}) for 
\(\bra{s}^{even}\) we obtain:
\begin{eqnarray}
\bra{s}\,\anni{b}{p'}\,\anni{b}{p}
\,\ket{\rho}^{even}&=&
\frac{2\,(1-\rho)^{L}}{1\,+ \,(1-2\rho)^{L}} 
\times \nonumber\\
 & &\bra{0}\prod_{p_{1}>0}(1+\cot\left(\,
\frac{\pi p_{1}}{L}\,\right)\anni{b}{p_{1}}
\anni{b}{-p_{1}})\,\anni{b}{p'}\,
\anni{b}{p}\prod_{p_{2}>0}
(1+\mu^{2}\cot\left(\,\frac{\pi p_{2}}{L}
\,\right)\crea{b}{-p_{2}}\crea{b}{p_{2}})
\ket{0}.
\label{35}
\end{eqnarray}
 It is easily seen that if \(p\not=-p'\)
 this is equal to
\begin{eqnarray}
&=&\frac{2\,(1-\rho)^{L}}{1\,+\,(1-2\rho)^{L}}\,
\prod_{p_{1}\not=p,p'}\left(1+
\mu^{2}\cot^{2}\,\left(\,\frac{\pi p_{1}}{L}\,\right)\,\right)
\nonumber\\
& & \mbox{}\times\,\bra{0}\,(1+\cot\,
\left(\,\frac{\pi p'}{L}\,\right)\,\anni{b}{p'}\,\anni{b}{-p'})
\,\anni{b}{p'}\,(1+\mu^{2}\cot\left(\,\frac{\pi p'}{L}\,
\right)\crea{b}{-p'}\,\crea{b}{p'})\,\ket{0}\nonumber\\
 & & \mbox{}\times\,\bra{0}\,(1+\cot\left(\,
\frac{\pi p}{L}\,\right)\,\anni{b}{p}\,\anni{b}{-p})\,
\anni{b}{p}\,(1+\mu^{2}\cot\left(\,\frac{\pi p}{L}\,
\right)\crea{b}{-p}\,\crea{b}{p})\,\ket{0}\nonumber\\
 &=&0.
\label{36}
\end{eqnarray}
 On the other hand if \(p=-p'\), we obtain
\begin{eqnarray}
 &=&\frac{2\,(1-\rho)^{L}}{1\,+ \,(1-2\rho)^{L}}\,
\prod_{p_{1}\not=p'}\left(1+\mu^{2}\cot^{2}
\,\left(\,\frac{\pi p_{1}}{L}\,\right)\,\right)
\nonumber\\
 & & \mbox{}\times\,\bra{0}\,(1+\cot\,\left(\,
\frac{\pi p'}{L}\,\right)\,\anni{b}{p'}\,\anni{b}{-p'})\,
\anni{b}{p'}\,\anni{b}{-p'}\,(1+\mu^{2}\cot\,
\left(\,\frac{\pi p'}{L}\,\right)\crea{b}{-p'}\,
\crea{b}{p'})\,\ket{0}\nonumber\\
 &=&\frac{\mu^{2}\,\cot\,(\,\frac{\pi p'}{L}\,)}{1+
\mu^{2}\cot^{2}\,(\,\frac{\pi p'}{L}\,)}\;.
\label{37}
\end{eqnarray}
 So we have
\beq
\bra{s}\,\anni{b}{p'}\,\anni{b}{p}\,\ket{\rho}^{even}
\;=\;\frac{\mu^{2}\,\cot\,(\,\frac{\pi p'}{L}\,)}{1
 +\mu^{2}\cot^{2}\,(\,\frac{\pi p'}{L}\,)}\,\delta_{p,-p'}.
\label{38}
\eeq

 If we insert this into (\ref{31}) and take the
 thermodynamic limit \(L\:\rightarrow\:\infty\),
 we get after a slight
 rearrangement of terms
\beq
\langle\noc{j}(t)\rangle\;=\;\frac{1}{2\pi}
\int_{-\pi}^{\pi}\,dp\;
 e^{-(\epsilon_{p}+\epsilon_{-p})t}\;
\frac{\rho^{2}}{\rho^{2}+(1-\rho)^{2}
\tan^{2}(\frac{p}{2})}
\label{39}
\eeq
 where \(\epsilon_{p}\;=\;1\;-\;\cos\,p
\;-\;i\,\eta\,\sin\,p\).

 Notice that due to the translational 
 invariance of the initial state
 there is no dependence on the site label \(j\)
 and, since \(\epsilon_{p}\)
 appears in combination with \(\epsilon_{-p}\)
 there is no dependence on the 
 anisotropy parameter \(\eta\) either
 ($\epsilon_{p}+\epsilon_{-p}\;=\;2\;-\;2\cos p $).

 Now we keep \(\rho\) finite but take 
\(t\,\rightarrow\,\infty\). Then, making
 the substitution \(p'\,=\,p\,\sqrt{t}\),
 we obtain by expanding the 
 trigonometric functions
\beq
\langle\noc{j}(t)\rangle\;=\;\frac{1}{2\pi\sqrt{t}}
\int_{-\infty}^{+\infty}
\,dp'\;e^{-p'^{2}}\;=\;
\frac{1}{2\sqrt{\pi t}}
\label{40}
\eeq
 which is the known result given
 in reference \cite{r7}.

 But also another scaling limit is of
 interest: We take 
\(t\,\rightarrow\,\infty\) and  
\(\rho\,\rightarrow\,0\) such that 
\(\rho^{2}\,t\:=\:const\). Then we have
\beq
\langle\noc{j}(t)\rangle\;=\;\frac{1}{2\pi\sqrt{t}}
\int_{-\infty}^{+\infty}\,dp'\;e^{-p'^{2}}\;
\frac{1}{1+\frac{p'^{2}}{4\rho^{2}t}}\;.
\label{41}
\eeq
 This integral can be calculated by
 the convolution theorem, giving
\beq
\langle\noc{j}(t)\rangle\;=\;\rho\,e^{4\rho^{2}t}\,
\mbox{Erfc}\,(\,2\rho\sqrt{t}\,)
\label{42}
\eeq
 a result also given in \cite{r7}.
 Notice that when \(\rho\sqrt{t}\,
\rightarrow\,\infty\) we recover 
\(\langle\noc{j}(t)\rangle\:\propto\:t^{-1/2}\).
 The thermodynamic limit (\ref{39}) also holds
 for the odd subsector, so that
 for the superposition of the two states, from 
 the even and the odd sector, 
 we obtain the same result. 

\section{The local average density for the step function state}
\label{sec5}

 In this section we calculate the time evolution 
 of the local average density
 for an initial state where the lattice
 is empty to the left of site $0$ and 
 full from there onwards, i.e. the initial
 local density is a step function. 
 The fermion representation of this state,
 which we will represent by 
\(\ket{\Phi}\), is, according to (\ref{28})
\beq
\ket{\Phi}\;=\;\prod_{k=0}^{L/2-1}\,\crea{a}{k}\ket{0}.
\label{43}
\eeq

 The lattice sites have been 
 renumbered from \(-L/2\)
 to \(L/2-1\), so that the step is located
 at \(0\), as stated above.
 Since this state is a product state in
 position space, it is convenient to 
 express \(\bra{s}\,\crea{b}{p}\,\anni{b}{p'}
 \ket{\Phi}\) in terms of 
 \(\bra{s}\,\crea{a}{k}\,\anni{a}{l}\ket{\Phi}\)
  with the aid of
 eq. (\ref{21}) - (\ref{22}). One gets
\beq
\bra{s}\,\crea{b}{p}\,\anni{b}{p'}\ket{\Phi}\:=\:
\frac{1}{L}\,\sum_{k,l=-L/2}^{L/2-1}\,
 e^{\frac{2\pi i}{L}(p'l-pk)}\,\bra{s}\,
\crea{a}{k}\,\anni{a}{l}\ket{\Phi}.
\label{44}
\eeq
 In the infinite volume limit,  
 $L\rightarrow\infty$, (\ref{32}) becomes,
 after the redefinition of (\ref{21}),
 (\ref{22}) such that the $\sqrt{L}$
 factor is absorbed into $\crea{b}{p}$,
 $\anni{b}{p'}$,
\beq
\langle\noc{j}(t)\rangle\;=\;\frac{1}{(2\pi)^2}
\,\int_{-\pi}^{\pi}\,dp\,
\int_{-\pi}^{\pi}\,dp'\,
 e^{i\,(p-p')j-(\epsilon_{-p}+\epsilon_{p'})t}
\,\bra{s}\,\crea{b}{p}\,
\anni{b}{p'}\,\ket{\Phi}.
\label{45}
\eeq
 Also, notice that we have absorbed the factors 
 $\frac{2\pi}{L}$ in the 
 redefinition of $p$ and $p'$, that are now
 continuous variables.
 Using (\ref{44}), we express $\bra{s}\,
 \crea{b}{p}\,\anni{b}{p'}\,\ket{\Phi}$,
 in terms of its Fourier transform $\bra{s}\,
 \crea{a}{k}\,\anni{a}{l}\,\ket{\Phi}$.
 One can then express
 this operators back into the spin language
 with the help of (\ref{16}) - (\ref{18}),
 obtaining
\begin{equation}
\bra{s}\,\crea{a}{k}\anni{a}{l}\,\ket{\Phi}=\left\{
\begin{array}{ll}
\bra{s}\,\spdo{k}\hat{\sigma}^{z}_{k+1}
\ldots\hat{\sigma}^{z}_{l-1}
\spup{l}\,\ket{\Phi}& \mbox{if $k<l$}\\
\\
 \bra{s}\,\spdo{k}\spup{k}\,\ket{\Phi}& 
\mbox{if $k=l$}\\
\\
\bra{s}\,\spup{l}\hat{\sigma}^{z}_{l+1}
\ldots\hat{\sigma}^{z}_{k-1}
\spdo{k}\,\ket{\Phi}& \mbox{if $k>l$}
\end{array}
\right. .
\label{46}
\end{equation}

 In this language, this expression is also
 easily computable for a half step profile, 
 i.e for an initial state in which the lattice
 is empty up to site zero
 and half full to the right of $0$
 (although the decomposition obtained
 in appendix B is not valid, 
 because this state does not have a simple
 representation in terms of fermion operators). 
 The results are
\begin{equation}
\bra{s}\,\crea{a}{k}\anni{a}{l}\,\ket{\Phi}=
\left\{
\begin{array}{ll}
\left\{
\begin {array}{ll}
 (-1)^{l} & \mbox{if $k<0$ and $l\geq 0$}\\
 0 & \mbox{otherwise}
\end{array}
\right.
 &\mbox{if $k<l$}\\
 1 & \mbox{if $k=l$ and $l\geq 0$}\\
 0 & \mbox{if $k>l$}
\end{array}
\right.
\label{47}
\end{equation}
 for the step function profile, and
\begin{equation}
\bra{s}\,\crea{a}{k}\anni{a}{l}\,
\ket{\Phi_{1/2}}=
\left\{
\begin{array}{ll}
\left\{
\begin {array}{ll}
 1/4 & \mbox{if $l\geq 1$
 and $k=l-1$}\\
 1/2 & \mbox{$l=0$ and $k<0$}\\
 0 & \mbox{otherwise}
\end{array}
\right.
 &\mbox{if $k<l$}\\
 1/2 & \mbox{if $k=l$ and $l\geq 0$}\\
\left\{
\begin{array}{ll}
 1/4 & \mbox{$k\geq 1$ and $l=k-1$}\\
 0 & \mbox{otherwise}
\end{array}
\right.
 & \mbox{if $k>l$}
\end{array}
\right.
\label{48}
\end{equation}
 for the half step profile.

 Inserting these two expressions
 in (\ref{44}), one obtains
\beq
\bra{s}\,\crea{b}{p}\,\anni{b}{p'}\ket{\Phi}\:=\:
\sum_{l=0}^{\infty}\,e^{i(p'-p)l}\;+\;
\sum_{l=0}^{\infty}\,(-1)^{l}\,e^{ip'l}\,\sum_{k=1}^{\infty}\,
 e^{ipk}
\label{49}
\end{equation}
 for the step function profile, and
\beq
\bra{s}\,\crea{b}{p}\,\anni{b}{p'}
\ket{\Phi_{1/2}}\:=\:\frac{2+e^{ip'}+
 e^{-ip}}{4}\,
\sum_{l=0}^{\infty}\,e^{i(p'-p)l}\;+\;\frac{1}{2}
\,\sum_{k=1}^{\infty}\,
 e^{ipk}
\label{50}
\end{equation}
 for the half step profile. One has to insert
 appropriate imaginary parts in
 $p$ and $p'$ to insure the convergence
 of the geometric series.

 It is easy to see that these expressions give
 the correct boundary conditions
 at $t=0$: if one inserts (\ref{49}) into
 (\ref{45}), the first term of the 
 r.h.s. just gives the step function
 when the integrations are 
 performed over $p$ and $p'$.
 The second term is equal to 
 $\sum_{l=0}^{\infty}\,(-1)^{l}\,
 \delta_{j,l}\,\sum_{k=1}^{\infty}\,
 \delta_{-j,k}$, which is identically zero. 
 A similar result holds for (\ref{50}).

 At finite time, one obtains from (\ref{45})
 and (\ref{49}) or (\ref{50})
 respectively, the exact time evolution of
 the density profile in terms of 
 infinite sums of products of two modified
 Bessel functions. More interesting,
 however, is the long time scaling limit,
 $t\rightarrow\infty$.
One performs a small momentum expansion of
the trigonometric functions. The two 
expressions become identical, and we get
\beq
\langle\noc{j}(t)\rangle\;=\;\frac{1}{(2\pi)^2t}\,
\int_{-\infty}^{\infty}\,dp\,
\int_{-\infty}^{\infty}\,dp'\,
 e^{i\,(p-p')y-(p^{2}+p'^{2})}\,
\left\{\sum_{l=0}^{\infty}\,
 e^{i(p'-p)\tilde{l}}\;+\;
\frac{1}{2}\sum_{k=1}^{\infty}\,
 e^{ip\tilde{k}}\right\}
\label{51}
\eeq
 where we have performed the substitutions 
 $p\rightarrow p\sqrt{t}$ and
 $p'\rightarrow p'\sqrt{t}$, and accordingly, 
 $y=\frac{j-\eta t}{\sqrt{t}}$,
 $\tilde{l}=\frac{l}{\sqrt{t}}$ and 
 $\tilde{k}=\frac{k}{\sqrt{t}}$. The integrals 
 can then be performed,
 giving
\beq
\langle\noc{j}(t)\rangle\;=\;\frac{1}{(2\pi)t}\,
\left\{\sum_{l=0}^{\infty}\,e^{-(y-\tilde{l})^2}\;+\;
\frac{1}{2}\,e^{-y^{2}/2}\,\sum_{k=1}^{\infty}\,
 e^{-\frac{(y+\tilde{k})^{2}}{2}}\right\}.
\label{52}
\eeq
 We can convert the sums into integrals if
 we absorb a factor of $t^{-1/2}$.
 The final result is 
\beq
\langle\noc{j}(t)\rangle\;=\;\frac{1}{4\sqrt{\pi t}}\,
\left\{1\;+\;\mbox{Erf}(y)\;+\;\frac{e^{-y^{2}/2}}{\sqrt{2}}\,
\left[\,1-\mbox{Erf}\left(\frac{y}{\sqrt{2}}\right)
\,\right]\right\}
\label{53}
\eeq
 where $\mbox{Erf}(x)=\frac{2}{\sqrt{\pi}}
\,\int_{0}^{x}\,dt\,
 e^{-t^{2}}$. 
 A plot of (\ref{53}) is given in Figure~1.
 Notice that the density for 
 large times depends on $\eta$ only through
 the scaling variable $y$. This means that the
 effect of the driving field
 can be completely absorbed by making a
 Galilean transformation to a
 frame moving with velocity $\eta$.
 This agrees with the general result
 obtained in \cite{r5}.
 Also, one can easily see that this
 expression has the right behaviour when
 $y\rightarrow\pm\infty$. Indeed, when
 $y\rightarrow+\infty$ it agrees
 with the limit (\ref{40}). When
 $y\rightarrow-\infty$ it goes to zero
 as it should be. The diffusion front widens as
 $t^{-1/2}$, but, as opposed to pure diffusion,
 the value of the density at 
 $y=0$   is not the turning point of the
 profile function with half of the 
 value  at $y=\infty$. Instead, the turning
 point lies at a distance proportional
 to $\sqrt{t}$ in the initially empty region $y<0$.

 This behaviour can be understood
 qualitatively through the following mean 
 field argument, where $n_{0} (t)$  is the
 density at site $0$ and 
 $n_{\infty}(t)$ the density very far to the
 right of zero. Take, for simplicity
 $\eta=0$ and consider early times such that
 the density to the left of zero is 
 very small and to the right is very close to $1$.
 Then the mean field equation
 of motion for $n_{0}$ simplifies to:
 $\dot{n}_{0}(t)\;=\;
 -\frac{1}{2} n_{0}(t)\;-\;\frac{1}{2} n_{0}^{2}(t)$
 (the second term of the r. h. s. 
 is absent in the pure diffusive case). 
 On the other hand the mean field equation 
 for $n_{\infty}(t)$ is $\dot{n}_{\infty}(t)
 \;=\;-n_{\infty}^{2}(t)$. In the pure 
 diffusive case the r.h.s. is of course zero.
 So the rate of decay of $n_{\infty}(0)$ 
 is comparable to the rate of decay of
 $n_{0}(0)$, as opposed to 
 the pure diffusive case in which this rate
 is zero. So, one should expect that, 
 at large times, $n_{0}(t)>\frac{1}{2}n_{\infty}(t)$.
 That is
 in fact what happens. Also, notice that
 the result obtained
 in the asymptotic large time limit is
 independent of whether we consider
 a step function profile or a half step
 function one. So we expect this result 
 to hold for any step function with non-zero
 initial density to the right of 
 site zero.

 Following the same mean field arguments,
 we should expect that if the rate of 
 annihilation  $\lambda$ is larger than
 $1$ ($\lambda=1$ being the
 free fermion condition) then one  would
 see a (transient) bump  developing at
 $y=0$, which would result from the fact
 that the rate for an event
 to the right of zero would be larger than
 the rate of an event at 
 site $0$ at early times. It would be
 interesting to verify
 this conjecture in Monte Carlo simulations.

\section{Application to experiments}
\label{sec6}

 As suggested in \cite{r10,r11} we would like to
 compare our exact theoretical
 results for the annihilation process to the
 experimental results mentioned in 
 the introduction by modeling the excitons dynamics
 by the coagulation process
 on the lattice. We have to make use of a
 similarity transformation
 to relate the two processes. Following
 \cite{r10} we define a new time 
 evolution operator by
\beq
\tilde{H}\;\equiv\;B\,H\,B^{-1}
\label{54}
\eeq
 where $B$ is assumed to have the form
\beq
 B\;=\;{\cal{B}}^{\otimes L}
\label{55}
\eeq
 and the matrix $\cal{B}$ acts on a single site.
 Since we are studying two state 
 systems $\cal{B}$ will be a $2\times2$ matrix.

 The diffusion-annihilation process which we
 are considering involves only
 nearest neighbour interactions. Hence
 $H$ can be written as
\beq
 H\;=\;\sum_{i}\,H_{i,i+1}
\label{56}
\eeq
 as can be easily seen from equation (\ref{13}).
 We now investigate the relation 
 between (\ref{56}) and the Hamiltonian describing
 the diffusion-coagulation process
\beq
 A \: \emptyset \: \longrightarrow \: 
\emptyset \: A \;\;\;\;\; \tilde{D}_{R}
\label{57}
\eeq
\beq
 \emptyset \: A \: \longrightarrow \: A 
\: \emptyset \;\;\;\;\; \tilde{D}_{L}
\label{58}
\eeq
\beq
 A \: A \: \longrightarrow \: \emptyset 
\: A  \;\;\;\;\; \tilde{\gamma}_{R} 
\label{59}
\eeq
\beq
 A \: A \: \longrightarrow \:A\: \emptyset  
\;\;\;\;\; \tilde{\gamma}_{L}
\label{60}
\eeq
 with (in general anisotropic) diffusion
 rates $\tilde{D}_{R,L}$ and 
 coagulation rates $\tilde{\gamma}_{R,L}$.

 Again $\tilde{H}$ is of the form (\ref{56}).
 A particularly convenient way to 
 represent $H_{i,i+1}$ and $\tilde{H}_{i,i+1}$
 is in a matrix form 
 involving the two site states
 (there are $2\times 2\:=\:4$ states,
 hence $H_{i,i+1}$, $\tilde{H}_{i,i+1}$ 
 are $4\times 4$
 matrices)
\beq
 H_{i,i+1}\;=\;\left(
\begin{array}{cccc}
 0 & 0 & 0 & \!\!\!\!-\lambda\\
 0 &\frac{1-\eta}{2}&\!\!\!\!-\frac{1+\eta}{2}&0\\
 0 &\!\!\!\!-\frac{1-\eta}{2}&\frac{1+\eta}{2}&0\\
 0 & 0 & 0 & \lambda
\end{array}\right)
\label{61}
\eeq
\beq
\tilde{H}_{i,i+1}\;=\;\left(
\begin{array}{cccc}
 0 & 0 & 0 & 0\\
 0 &\tilde{D}_{L}&\!\!\!\!-\tilde{D}_{R}&
\!\!\!\!-\tilde{\gamma}_{R}\\
 0&\!\!\!\!-\tilde{D}_{L}&\tilde{D}_{R} &
\!\!\!\!-\tilde{\gamma}_{L}\\
 0 & 0 & 0 & \tilde{\gamma}_{R}+\tilde{\gamma}_{L}
\end{array}\right).
\label{62}
\eeq

 Since $H$ is of the form (\ref{56}) we
 only need to consider the
 effect of ${\cal{B}}\bigotimes{\cal{B}}$
 on (\ref{61}). It can be shown 
\cite{Krebs} that the choice
\beq
 {\cal{B}}\;=\;\left(
\begin{array}{cc}
 1&\!\!\!\! -1 \\
 0& 2 
\end{array}\right)
\label{63}
\eeq
 maps (\ref{61}) into the form (\ref{62}). The new
 parameters $\tilde{D}_{R}$, $\tilde{D}_{L}$,
 $\tilde{\gamma}_{R}$,
 $\tilde{\gamma}_{L}$ are given in terms of
 the old parameters $\lambda$,
 $\eta$ by
\beq
\tilde{D}_{R}\;=\;\frac{1}{2}\,(1+\eta)
\label{64}
\eeq
\beq
\tilde{D}_{L}\;=\;\frac{1}{2}\,(1-\eta)
\label{65}
\eeq
\beq
\tilde{\gamma}_{R}\;=\;\frac{1}{2}\,(\lambda+\eta)
\label{66}
\eeq
\beq
\tilde{\gamma}_{L}\;=\;\frac{1}{2}\,(\lambda-\eta).
\label{67}
\eeq

 Since in the exciton dynamics the rate of decay
 is very small compared to all
 other rates we can accept  equations (\ref{57}) -
 (\ref{60}) as a good 
 description of it.
 Also, as argued in \cite{r10}, in the
 real system $\tilde{D}_{R}\:=\:
\tilde{D}_{L}\:=\:\tilde{\gamma}_{R}
\:=\:\tilde{\gamma}_{L}$. From equations
 (\ref{64}) - (\ref{67}) it then 
 follows that $\eta=0$, $\lambda=1$.
 The first condition
 is just symmetric diffusion, but the second
 is the free fermion condition. 
 Hence, our previous discussion
 can be used to obtain quantities which are
 of experimental relevance. 
 However, we still have to relate the
 expectation values in the two problems.
 According to (\ref{54}) the 
 average value of some projection operator
 $F$, $\bra{s}\,F\,e^{-Ht}\,\ket{\Psi}$
 in the original problem is related to 
 the following quantity in the new
 problem
\begin{eqnarray}
\bra{s}\,F\,e^{-Ht}\,\ket{\Psi}&=&
\bra{s}\,F\,B^{-1}\,B\,e^{-Ht}\,B^{-1}
\,B\,\ket{\Psi}\nonumber\\
 &=&\bra{s}\,\tilde{F}\,e^{-\tilde{H}t}\,
\ket{\tilde{\Psi}}
\label{68}
\end{eqnarray}
 where $\tilde{F}\:=\:F\,B^{-1}$, 
 $\tilde{H}\:=\:B\,H\,B^{-1}$ and 
 $\ket{\tilde{\Psi}}\:=\:B\,\ket{\Psi}$.
 In particular, if $F\:=\:\noc{k}$
 and $\ket{\Psi}\:=\:\prod_{i}\,
 \left(\begin{array}{c}
 1-p_{i}\\p_{i}\end{array}\right)$ is a product
 state with average density
 $p_i$ on site $i$, then
\beq
\tilde{F}\;=\;\frac{1}{2}\,\noc{k}
\label{69}
\eeq
\beq
\ket{\tilde{\Psi}}\;=\;\prod_{i}\,\left(\begin{array}{c}
 1-2p_{i}\\2p_{i}\end{array}\right).
\label{70}
\eeq
 We see from (\ref{70}) that for $\ket{\tilde{\Psi}}$
 to be a physical state we
 must have $p_{i}\:\leq 1/2$ in the original problem.

 Let us apply (\ref{68}) - (\ref{70}) to the
 case when we have random initial
 conditions in the coagulation problem with
 an initial density $\rho$ (this corresponds
 to a density equal to $\frac{\rho}{2}$ in
 the annihilation problem).
 Then we get from expression (\ref{39})
 the following result for
 $\langle \noc{j}(t)\rangle_{coag}$
\beq
\langle\noc{j}(t)\rangle_{coag}\;=\;
\frac{1}{\pi}\int_{-\pi}^{\pi}\,dp\;e^{-2(1-\cos p)t}\;
\frac{\rho^{2}}{\rho^{2}+(2-\rho)^{2}
\tan^{2}(\frac{p}{2})}\;.
\label{71}
\eeq

 The scaling limit $\rho\rightarrow 0$, 
 $t\rightarrow \infty$, $\rho^{2}\,t\:=\:const$
(\ref{42}) translates to
\beq
\langle\noc{j}(t)\rangle_{coag}\;=\;
\rho\,e^{\rho^{2}t}\,\mbox{Erfc}\,(\,\rho\sqrt{t}\,).
\label{72}
\eeq
 This was experimentally verified within 5 \% \cite{r1}.

 The expression (\ref{72}) is already known
 from the literature \cite{r12}
 where it was obtained using a continuum
 approximation for the master equation.
 Our calculation provides a rigorous proof
 of the validity of this expression
 in the scaling regime.

 A similar mapping from the step function state
 of section~\ref{sec5} is inappropriate
 because this state maps into a non-physical
 state of the coagulation problem. 
 However, the half step function for the
 annihilation problem maps directly
 into the results for the step function profile
 in the coagulation problem
 and hence the appropriate version of
 $(\ref{53})$, namely
\beq
\langle\noc{j}(t)\rangle_{coag}\;=\;
\frac{1}{\sqrt{4\pi t}}\,
\left\{1\;+\;\mbox{Erf}(y)\;+\;
\frac{e^{-y^{2}/2}}{\sqrt{2}}\,
\left[\,1-\mbox{Erf}\left(\frac{y}{
\sqrt{2}}\right)\,\right]\right\}
\label{73}
\eeq
 can be verified experimentally. Since,
 as argued in the preceding section, 
 the initial density does not seem to play a
 role, the fact that it is derived
 for an initial density $\rho_{right}=1$
 is immaterial. (\ref{73}) should
 be correct for any step in the late time regime.

\section{Conclusions}

 We have computed time averages for some
 specific initial states in a
 simple reaction-diffusion system where
 exclusion particles hop with
 rates $(1\pm\eta)/2$ to the right and
 left respectively if the nearest sites 
 are empty, and which annihilate in pairs
 with rate one if the nearest-neighbour 
 site is occupied. We obtained the following results.

 (i) For states with random initial
 conditions we generalized
 earlier results \cite{r7,r13,r14} to all times
 and arbitrary densities. Also,
 we have obtained a scaling limit for large times and small
 initial densities. Further,
 we have shown that
 due to the fact that those states are
 product states in the fermion representation,
 a particular
 product decomposition holds for quantities
 which are useful for the
 the calculation of multi-time and
 higher order equal-time 
 correlation functions.

 (ii) For the step function and half step
 function state, we calculated the long
 time behaviour of the time-dependent
 average density. We
 concluded that the competition between
 the annihilation and diffusion process
 shifts the profile to the left of the moving front. 
 It would be interesting to check this
 prediction experimentally.
 A product decomposition
 holds for the step function state.

 (iii) The use of the similarity transformation
 has shown that the scaling
 limit obtained above of our exact
 lattice calculation reproduces the
 known results obtained in a continuum approach.

 It would be very interesting to see what
 would happen to the density
 profile in the step function states
 if one goes away from the value
 $\lambda=1$. The results of appendix B
 allow for a perturbative calculation 
 but to higher order in $\lambda-1$
 the procedure becomes 
 rather tedious. This could be a task that
 could be performed by simulations.

\noindent{
 {\bf Acknowledgments:}
 G.M.S. would like to thank the Institute
 for Advanced Study, Princeton
 where this work was completed for kind hospitality.
 J.E.S. is  supported by the  Grant:
 PRAXIS XXI/BD$/3733/94$ - JNICT -
 PORTUGAL. G.M.S. is supported by an EC fellowship
 under the Human Capital 
 and Mobility program.}

\appendix

\section*{Appendix A Calculation
 of multi-time correlation functions in states 
 with random initial conditions}
\label{apA}

 In section~\ref{sec3} we have indicated
 how the calculation 
 of a multi-time correlation
 function could be reduced to the
 calculation of quantities 
 like \(\bra{s}\,\anni{b}{p_{1}}
 \ldots\anni{b}{p_{2N}}\ket{\Psi}\).
 Here, we show how 
 they can be calculated 
 for a 
 \(\ket{\rho}^{even}\) state.

 Firstly, we define
\beq
 Z[\eta_{p},\eta_{-p}]\;=\;\bra{0}\,e^{\sum_{p>0}
 (\,\cot(\frac{\pi p}{L})\,\anni{b}{p}\,\anni{b}{-p}
 +\etal{p}\,\anni{b}{p}+\etal{-p}\,
\anni{b}{-p}\,)}
\,\ket{\rho}^{even}
\label{A1}
\eeq
 where the quantities \(\eta_{p}\), \(\eta_{-p}\)
 are Grassmann variables
 obeying \(\{\eta_{p},\eta_{p'}\}\:=\:0\),
 \(\{\eta_{p},\anni{b}{p'}\}\:=\:0\).
 Their presence makes all the terms in the
 exponential commute with each 
 other. Having this in mind and the fact
 that \(e^{\etal{p}\,\anni{b}{p}}\:=\:
 1\:+\:\eta_{p}\,\anni{b}{p}
 \;\; ((\eta_{p}\,\anni{b}{p})^{2}=0)\),
 we can write (\ref{A1}) as
 \beq
 Z[\eta_{p},\eta_{-p}]\;=\;\bra{0}\,
 e^{\sum_{p>0}\cot(\frac{\pi p}{L})\,
\anni{b}{p}\,\anni{b}{-p}}\,\prod_{p}\,
(\,1\;+\;\eta_{p}\,\anni{b}{p}\,)\,
\ket{\rho}^{even}.
\label{A2}
\eeq
 Notice that \(Z[\,0\,]\:=\:\bra{0}\,e^{\sum_{p>0}
\cot(\frac{\pi p}{L})\,
\anni{b}{p}\,\anni{b}{-p}}\,
\ket{\rho}^{even}\:=\:\langle\,s\,
\ket{\rho}^{even}\:=\:1\). Also, from 
 (\ref{A2}) and from the fact that 
\(\eta_{p}\) and \(\anni{b}{p'}\)
 anticommute one gets
\beq
\frac{\partial\,Z[\eta_{p},\eta_{-p}]}{
\partial\,\eta_{p_{1}}}\;=
\;\bra{0}\,e^{\sum_{p>0}\cot(
\frac{\pi p}{L})\,\anni{b}{p}\,
\anni{b}{-p}}\,\prod_{p}\,
 (\,1\;+\;\eta_{p}\,\anni{b}{p}\,)\,\anni{b}{p_{1}}
\,\ket{\rho}^{even}
\label{A3}
\eeq
 where the product over \(p\) still contains
 \(p_{1}\) since \(\anni{b}{p}^{2}=0\).

 One can go on differentiating with
 respect to the \(\eta_{p}\)'s 
(an even number of times). Setting
 \(\eta_{p}=0\) one gets
\begin{eqnarray}
\partial_{\eta_{p_{1}}}\ldots\partial_{\eta_{p_{2N}}}\,Z[\eta_{p},
\eta_{-p}]\mid_{\eta_{p}=0}
 &=&\bra{0}\,e^{\sum_{p>0}\cot(\frac{\pi p}{L})\,
\anni{b}{p}\,\anni{b}{-p}}\,\anni{b}{p_{1}}
\ldots\anni{b}{p_{2N}}\,
\ket{\rho}^{even}\nonumber\\
 &=& \bra{s}\,\anni{b}{p_{1}}\ldots\anni{b}{p_{2N}}
\ket{\rho}^{even}.
\label{A4}
\end{eqnarray}
 Hence \(Z[\eta_{p},\eta_{-p}]\) is the
 generating function for the quantities 
\(\bra{s}\,\anni{b}{p_{1}}\ldots\anni{b}{p_{2N}}\ket{\rho}^{even}\).
 As it can be seen from the  preceding steps,
 (\ref{A4})
 does not depend on \(\ket{\rho}^{even}\).
 It is true for any state in
 the even sector. But the form (\ref{34})
 for \(\ket{\rho}^{even}\)
 allows us to calculate \(Z[\eta_{p},\eta_{-p}]\)
 explicitly. In fact
\begin{eqnarray}
 Z[\eta_{p},\eta_{-p}]&=&{\cal N}\,\bra{0}\,
 e^{\sum_{p>0}(\,\cot(\frac{\pi p}{L})\,\anni{b}{p}
\,\anni{b}{-p}+\etal{p}\,\anni{b}{p}+\etal{-p}\,
\anni{b}{-p}\,)}
\,\prod_{p'>0}\,
 (1\:+\:\mu^{2}\,\cot\,\left(\,\frac{\pi p'}{L}\,\right)
\,\crea{b}{-p'}\,\crea{b}{p'}\,)\,\ket{0}
\nonumber\\
 &=&{\cal N}\,\bra{0}\,\prod_{p>0}(1+
\eta_{p}\anni{b}{p}+\eta_{-p}\anni{b}{-p}+
(\,\cot\,\left(\,\frac{\pi p}{L}\,\right)+
\eta_{-p}\eta_{p}\,)\anni{b}{p}\,
\anni{b}{-p})\nonumber\\
 & &\mbox{}\times\prod_{p'>0}\,
 (1\:+\:\mu^{2}\,\cot\,\left(\,\frac{\pi p'}{L}\,
\right)\,\crea{b}{-p'}\,
\crea{b}{p'}\,)\,\ket{0}\nonumber\\
 &=&{\cal N}\,\prod_{p>0}\left[1+\mu^{2}
\cot\left(\,\frac{\pi p}{L}\,\right)
\,(\,\cot\left(\,\frac{\pi p}{L}\,\right)+
\eta_{-p}\eta_{p})\,\right]
\label{A5}
\end{eqnarray}
 where \({\cal N}\) is the normalization
 factor of \(\ket{\rho}^{even}\) and
 the last step of (\ref{A5}) is similar to the
 manipulations done in equations 
 (\ref{35}) - (\ref{37}). Also from 
\(Z[\,0\,]\:=\:1\) it follows that 
\({\cal N}\:=\:[\,\prod_{p>0}\,(1+\mu^{2}
\cot^{2}(\,\frac{\pi p}{L}\,)\,)\,]^{-1}\).
 If we define \(W\:=\:\ln\,
 Z[\eta_{p},\eta_{-p}]\), we get for
\(W[\eta_{p},\eta_{-p}]\)
\begin{eqnarray}
 W[\eta_{p},\eta_{-p}]&=&\sum_{p>0}\,
 \{\:\ln\,[\,1\;+\;\mu^{2}\,
 \cot\left(\,\frac{\pi p}{L}\,\right)\,
 (\,\cot\left(\,\frac{\pi p}{L}\,\right)+
\eta_{-p}\eta_{p}\,)\,]\:-\;\ln\,[\,1\;+\;\mu^{2}\,
\cot^{2}\left(\,\frac{\pi p}{L}\,\right)\,]\:\}
\nonumber\\
 &=&\sum_{p>0}\,\ln\,\left(\,1\:+\:
\frac{\mu^{2}\,\cot(\,\frac{\pi p}{L}\,)}{1+\mu^{2}
\,\cot^{2}(\,\frac{\pi p}{L}\,)}\,
\eta_{-p}\,\eta_{p}\,\right)
\nonumber\\
 &=&\sum_{p>0}\,\frac{\mu^{2}\,\cot(\,
\frac{\pi p}{L}\,)}{1+\mu^{2}\,\cot^{2}(
\,\frac{\pi p}{L}\,)}\,
\eta_{-p}\,\eta_{p}
\label{A6}
\end{eqnarray}
 since \((\eta_{-p}\,\eta_{p})^{2}\:=
\:(\eta_{-p}\,\eta_{p})^{3}
\:=\:\ldots\:=\:0\).

 It is known from field theory that \(W[\eta_{p},\eta_{-p}]\)
 is the generating 
 function for the connected part of the quantities
 \(\bra{s}\,\anni{b}{p_{1}}
 \ldots\anni{b}{p_{2N}}\ket{\rho}^{even}\). 
 Since \(W\) is quadratic all the
 connected parts vanish for \(N>1\) and
 Wick's theorem holds. From (\ref{A6})
 one concludes
\beq
\frac{\partial\,W}{\partial\,\eta_{p}}\mid_{\eta_{p}=0}\;=\;0.
\label{A7}
\eeq
 Also
\begin{eqnarray}
\bra{s}\,\anni{b}{p'}\,\anni{b}{p}\ket{\rho}^{even}&=&
\frac{\partial^{2}Z}{\partial\,\eta_{p'}\,\partial\,
\eta_{p}}\mid_{\eta_{p}=0}\:=\:
\left(\frac{\partial^{2}W}{\partial\,\eta_{p'}
 \,\partial\,\eta_{p}}
 +\frac{\partial\,W}{\partial\,\eta_{p'}}\,
\frac{\partial\,W}{\partial\,\eta_{p}}
\right)\mid_{\eta_{p}=0}
\nonumber\\
 &=&\frac{\partial^{2}W}{\partial\,\eta_{p'}
 \,\partial\,\eta_{p}}\mid_{\eta_{p}=0}\:=\:
 \frac{\mu^{2}\,\cot\,(\,\frac{\pi p'}{L}\,)}{
 1+\mu^{2}\cot^{2}\,(\,
 \frac{\pi p'}{L}\,)}\,\delta_{p,-p'}
\label{A8}
\end{eqnarray}
 which agrees with (\ref{38}).

 Further, from Wick's theorem one has
\begin{eqnarray}
\bra{s}\,\anni{b}{p_{1}}\,\anni{b}{p_{2}}\,
\anni{b}{p_{3}}\,\anni{b}{p_{4}}\,
\ket{\rho}^{even}&=&\bra{s}\,\anni{b}{p_{1}}\,
\anni{b}{p_{2}}\,\ket{\rho}^{even}\,
\bra{s}\,\anni{b}{p_{3}}\,\anni{b}{p_{4}}\,
\ket{\rho}^{even}
\nonumber\\
 & &\mbox{}-\bra{s}\,\anni{b}{p_{1}}\,
\anni{b}{p_{3}}\,\ket{\rho}^{even}\,
\bra{s}\,\anni{b}{p_{2}}\,\anni{b}{p_{4}}\,
\ket{\rho}^{even}\nonumber\\
 & &\mbox{}+\bra{s}\,\anni{b}{p_{1}}\,
\anni{b}{p_{4}}\,\ket{\rho}^{even}\,
\bra{s}\,\anni{b}{p_{2}}\,
\anni{b}{p_{3}}\ket{\rho}^{even}.
\label{A9}
\end{eqnarray}
 And so on. Note that as already remarked above,
 (\ref{A4}) only holds
 for an even number of \anni{b}{p} operators.
 But all the correlation functions
 can be reduced to sums of quantities like
 \(\bra{s}\,\anni{b}{p_{1}}\ldots
 \anni{b}{p_{2N}}\ket{\rho}^{even}\)
 as discussed in section~\ref{sec3}.

 The equality (\ref{A9}) can be used to
 calculate \(\bra{s}\,\noc{j_{1}}(t_{1})
 \,\noc{j_{2}}(t_{2})\,
\ket{\rho}^{even}\). But the manipulations
 necessary, although straightforward, 
 become rather involved. A simplification
 takes place for \(\rho=1\) and 
\(\rho=1/2\), for in these cases 
\(\bra{s}\,\anni{b}{p'}\,\anni{b}{p}
\ket{\rho}\) becomes:

 For \(\rho=1\)
\beq
\bra{s}\,\anni{b}{p'}\,\anni{b}{p}
\ket{\rho}\;=\;\delta_{p,-p'};
\label{A10}
\eeq

 For \(\rho=1/2\)
\beq
\bra{s}\,\anni{b}{p'}\,\anni{b}{p}\ket{\rho}\;=
\;\frac{1}{2}\sin
\left(\,\frac{2\pi p'}{L}\,\right)\,\delta_{p,-p'}.
\label{A11}
\eeq

\section*{Appendix B Higher order correlation
 functions in the step function state}
\label{apB}

 In this appendix we develop a procedure for
 calculating higher order correlation 
 functions in the step function state $\ket{\Phi}$
 defined in section~\ref{sec5}. 
 The quantities which we wish to compute are 
 $\bra{s}\,\anni{a}{j_{1}}\ldots\anni{a}{j_{2N}}
 \,\ket{\Phi}$. Given that 
 $\ket{\Phi}$ is a product state these
 quantities are easily computed
 by expressing the annihilation operators
 in terms of the original spin operators
 (\ref{17}) - (\ref{18}) if $j_{1}<\ldots<j_{2N}$. 
 However, the quantities necessary for
 the calculation of correlation functions 
 are $\bra{s}\,\anni{b}{p_{1}}\ldots\anni{b}{p_{2N}}\,
\ket{\Phi}$. Expressing 
 these in terms of $\bra{s}\,\anni{a}{j_{1}}\ldots
\anni{a}{j_{2N}}\,\ket{\Phi}$ 
 by Fourier transformation involves all 
 possible orderings of $j_{1},\ldots 
 ,j_{2N}$ and the reordering of these terms
 becomes rather cumbersome. The 
 procedure that we develop avoids that.
 We start by introducing the coherent state
\beq
\ket{\eta}\:=\:e^{-\sum_{k}\etal{k}\crea{a}{k}}\,\ket{0}
\label{B1}
\eeq
 where the $\eta_{k}$'s are a set of Grassmann
 variables that anticommute with the 
 $\anni{a}{j}$, $\crea{a}{j}$. Developing
 (\ref{B1}) in powers of the $\eta_{k}$'s gives
\begin{eqnarray}
\ket{\eta}&=&\ket{0}\;-\;\sum_{j_{1}}\,\eta_{j_{1}}
\,\crea{a}{j_{1}}\,\ket{0}\;+\;
\sum_{j_{1}<j_{2}}\,\eta_{j_{2}}\,
\eta_{j_{1}}\,\crea{a}{j_{1}}\,\crea{a}{j_{2}}\,
\ket{0}\nonumber\\
 & &\mbox{}+\ldots+\;\sum_{j_{1}<j_{2}<\ldots<j_{L/2}}
\,\eta_{j_{L/2}}\ldots
\eta_{j_{1}}\,
\crea{a}{j_{1}}\ldots\crea{a}{j_{L/2}}\,
\ket{0}\nonumber\\
 & &\mbox{}+\ldots +\; \eta_{L/2-1}
\ldots\eta_{-L/2}\, \crea{a}{-L/2}
\dots\crea{a}{L/2-1}\,\ket{0}.
\label{B2}
\end{eqnarray}

 The state $\ket{\Phi}$ can be generated by
 taking derivatives with respect to 
 $\eta_{L/2-1},\ldots,\eta_{0}$ and then
 setting all the other $\eta$'s to 
 zero, i.e.
\beq
\ket{\Phi}\;=\;[\,\partial_{\eta_{0}}\ldots\partial_{\eta_{L/2-1}}\,
\ket{\eta}\,]_{\mbox{
 all other $\eta$'s $=0$}}\;=\;\left\{
\int\,d\eta_{0}\ldots d\eta_{L/2-1}\,\ket{\eta}
\right\}_{\mbox{all other $\eta$'s $=0$}}.
\label{B3}
\eeq
 The last step follows from the fact
 that there is no distinction 
 between integration and differentiation
 in Grassmann algebras. Now, in order
 to calculate $\bra{s}\,
\anni{a}{j_{1}}\ldots\anni{a}{j_{2N}}\,\ket{\Phi}$
 we define an enlarged 
$\ket{\Phi,\bar{\zeta}}$ state
\beq
\ket{\Phi,\bar{\zeta}}\;=\;\left\{
\int\,d\eta_{0}\ldots d\eta_{L/2-1}\,e^{-\sum_{j}\,\eta_{j}
\bar{\zeta}_{j}}\,\ket{\eta}
\right\}_{\mbox{all other $\eta$'s $=0$}}
\label{B4}
\eeq
 where $\zeta_{j}$ is another set of
 Grassmann variables anticommuting 
 both with $\eta_{j}$ 
 and $\anni{a}{j}$, $\crea{a}{j}$.
 Using $\anni{a}{j}\,\ket{\eta}\:=\:
 \eta_{j}\,\ket{\eta}$ gives
\begin{eqnarray}
\bra{s^{even}}\,\anni{a}{j_{1}}\ldots
\anni{a}{j_{2N}}\,\ket{\Phi,\bar{\zeta}}
 &=&
\left\{
\int\,d\eta_{0}\ldots d\eta_{L/2-1}\,
 e^{-\sum_{j}\,\eta_{j}
\bar{\zeta}_{j}}\,\eta_{j_{1}}\ldots
\eta_{j_{2N}}\,\langle s^{even}\ket{\eta}
\right\}_{\tilde{0}}
\nonumber\\
 &=&\partial_{\bar{\zeta}_{j_{1}}}\ldots
\partial_{\bar{\zeta}_{j_{2N}}}\,
\left\{
\int\,d\eta_{0}\ldots d\eta_{L/2-1}\,
 e^{-\sum_{j}\,\eta_{j}\bar{\zeta}_{j}}
\,\langle s^{even}\ket{\eta}\right\}_{\tilde{0}}
\label{B5}
\end{eqnarray}
 where the subscript $\tilde{0}$ is a
 reminder that we have to take the 
 remaining $\eta$'s equal
 to zero. Setting the $\bar{\zeta}$'s to
 zero gives the desired quantity 
 $\bra{s}\,\anni{a}{j_{1}}\ldots\anni{a}{j_{2N}}\,
\ket{\Phi}$ (remember that
 $\bra{s^{odd}}\,\anni{a}{j_{1}}\ldots\anni{a}{j_{2N}}
\,\ket{\Phi}$ is zero since
 $\ket{\Phi}\in$ even sector). Notice that we have
 not made any assumption on the
 order of the $j$'s in equation (\ref{B5}), which may be any,
 as already stressed 
 above. From (\ref{B5}), we conclude that 
\beq
 Z\,[\,\bar{\zeta}_{j}\,]\;=\;\left\{
\int\,d\eta_{0}\ldots d\eta_{L/2-1}\,e^{-\sum_{j}\,
\eta_{j}\bar{\zeta}_{j}}\,
\langle s^{even}\ket{\eta}\right\}_{\tilde{0}}
\label{B6}
\eeq
 is therefore a generating function for $\bra{s}\,
\anni{a}{j_{1}}\ldots
\anni{a}{j_{2N}}\,\ket{\Phi}$.

 Due to the form of \bra{s^{even}}, 
 the quantity $\langle s^{even}\ket{\eta}$ 
 can be easily calculated.
 We have
\begin{eqnarray}
\langle s^{even}\ket{\eta}&=&\bra{0}\,\prod_{p>0}\,(
1+\cot\left(\frac{\pi p}{L}\right)\,
\anni{b}{p}\,\anni{b}{-p}\,)\,\ket{\eta}
\nonumber\\
&=&\prod_{p>0}\,(1+\cot\left(\frac{\pi p}{L}
\right)\,\eta_{p}\,\eta_{-p}\,)
\;=\;e^{\sum_{p>0}\cot\left(\frac{\pi p}{L}
\right)\,\eta_{p}\,\eta_{-p}}
\label{B7}
\end{eqnarray}
 since  $\anni{b}{p}\,\ket{\eta}\:=\:
\eta_{p}\,\ket{\eta}$, where $\eta_{p}$ is the Fourier
 transform of $\eta_{k}$. Substituting
 this in (\ref{B6}) we get
\begin{eqnarray}
 Z\,[\,\bar{\zeta}_{j}\,]&=&\left\{
\int\,d\eta_{0}\ldots d\eta_{L/2-1}\,e^{-\sum_{j}\,\eta_{j}
\bar{\zeta}_{j}}\,
 e^{\frac{1}{2}\sum_{p}\cot\left(\frac{\pi p}{L}\right)\,
\eta_{p}\,\eta_{-p}}\right\}_{\tilde{0}}\nonumber\\
 &=&\left\{
\int\,d\eta_{0}\ldots d\eta_{L/2-1}\,e^{-\sum_{j}
\,\eta_{j}\bar{\zeta}_{j}}\,
 e^{\frac{1}{2}\sum_{k,l}\,F_{kl}\,\eta_{k}\,
\eta_{l}}\right\}_{\tilde{0}}
\label{B8}
\end{eqnarray}
 where $F_{kl}$ is the matrix
\beq
 F_{kl}\;=\;\frac{1}{L}\,\sum_{p}\,
\cot\left(\frac{\pi p}{L}\right)\,
\sin\left[\frac{2\pi}{L}p(k-l)\right].
\label{B9}
\eeq

 The last step of (\ref{B8}) follows by expressing 
$\eta_{p}$, $\eta_{-p}$ in terms of 
$\eta_{k}$, $\eta_{l}$ and using the Grassmann 
 character of the $\eta$'s.
 Setting to zero all $\eta_{j}$'s in which $j<0$ gives
\beq
 Z\,[\,\bar{\zeta}_{j}\,]\;=\;\int\,d\eta_{0}\ldots 
 d\eta_{L/2-1}\,e^{-\sum_{j>0}\,
\eta_{j}\bar{\zeta}_{j}}\,
 e^{\frac{1}{2}\sum_{k,l>0}\,F_{kl}\,\eta_{k}\,\eta_{l}}.
\label{B10}
\eeq
 This expression only involves the matrix elements
 of $F$ such that $k,l\geq 0$, and
 since $\bra{s}\,\anni{a}{j_{1}}\ldots\anni{a}{j_{2N}}
\,\ket{\Phi}$ involves the
 derivatives of $Z$ with respect to each $j_{i}$
 $i=1,\ldots,2N$, it will be zero
 if any of the $j_{i}$'s is less than zero.

 Now equation (\ref{B10}) is easily computed
 by completing the square  in the exponential
 and then shifting the integration variables.
 This gives
\beq
 Z\,[\,\bar{\zeta}_{j}\,]\;=\;e^{\frac{1}{2}
\sum_{k,l}\,\tilde{F}^{-1}_{kl}\,
\bar{\zeta}_{k}\,\bar{\zeta}_{l}}\,
\int\,[\,d\eta\,]\,e^{\frac{1}{2}
\sum_{k,l}\,\tilde{F}_{kl}\,\eta_{k}\,
\eta_{l}}
\label{B11}
\eeq
 where $\tilde{F}_{kl}$ is the matrix obtained
 by truncating $F_{kl}$ keeping only the
 elements
 $k,l\geq 0$ and $\tilde{F}^{-1}_{kl}$
 its inverse. The integral is 
 standard \cite{r15}, giving
\beq
 Z\,[\,\bar{\zeta}_{j}\,]\;=\;\sqrt{\mbox{det} 
\tilde{F}}\,e^{\frac{1}{2}\sum_{k,l}\,
\tilde{F}^{-1}_{kl}\,\bar{\zeta}_{k}\,
\bar{\zeta}_{l}}.
\label{B12}
\eeq

It is clear that $\sqrt{\mbox{det} 
\tilde{F}}\:=\:1$ since 
$Z[\,0\,]\:=\:\sqrt{\mbox{det} 
\tilde{F}}\:=\:\langle s\,\ket{\Phi}=1$.
 To obtain the explicit form of 
 $\tilde{F}^{-1}_{kl}$ we note that $F_{kl}$ is
 given by
\beq
 F_{kl}\;=\;\theta(k-l)\;-\;\theta(l-k)
\label{B13}
\eeq
 as can be seen by Fourier transforming it,
 obtaining (\ref{B9}). For an
 $L$-site system (\ref{B13}) is an
 $L\times L$ antisymmetric matrix with every
 matrix element equal to minus or plus 1
 depending on whether it is above or 
 below the diagonal. The matrix $\tilde{F}_{kl}$
 corresponds to the lower right 
 quadrant of $F_{kl}$ (obtained by suppressing
 the first half of the rows and 
 columns of $F$). Hence, apart from having
 different dimensions, $F$ and 
 $\tilde{F}$ have the same simple form.
 This can be used to confirm explicitly 
 that $\mbox{det} \tilde{F}\:=\;1$. Also,
 from the known Fourier transform of 
 $F$, or by direct methods, the inverse
 matrix $\tilde{F}^{-1}$ can be shown to 
 be
\beq
\tilde{F}_{kl}^{-1}\;=\;(-1)^{k-l}\,
[\,\theta(k-l)\;-\;\theta(l-k)\,]
\;\;\;\;k,l\geq 0.
\label{B14}
\eeq
 So, we finally obtain from equation (\ref{B12})
\beq
 Z\,[\,\bar{\zeta}_{j}\,]\;=\;e^{\frac{1}{2}\sum_{k,l}
\,\tilde{F}^{-1}_{kl}\,
\bar{\zeta}_{k}\,\bar{\zeta}_{l}}.
\label{B15}
\eeq

 As in the previous appendix this immediately
 shows that a form of Wick's 
 theorem holds for the quantities 
 $\bra{s}\,\anni{a}{j_{1}}\ldots\anni{a}{j_{2N}}
\,\ket{\Phi}$. For example
\begin{eqnarray}
\bra{s}\,\anni{a}{j_{1}}\,\anni{a}{j_{2}}\,
\anni{a}{j_{3}}\,\anni{a}{j_{4}}\,
\ket{\Phi}&=&\bra{s}\,\anni{a}{j_{1}}
\,\anni{a}{j_{2}}\,\ket{\Phi}\,
\bra{s}\,\anni{a}{j_{3}}\,\anni{a}{j_{4}}\,\ket{\Phi}
\nonumber\\
 & &\mbox{}-\bra{s}\,\anni{a}{j_{1}}
\,\anni{a}{j_{3}}\,\ket{\Phi}
\bra{s}\,\anni{a}{j_{2}}\,\anni{a}{j_{4}}
\,\ket{\Phi}\nonumber\\
 & &\mbox{}+\bra{s}\,\anni{a}{j_{1}}\,
\anni{a}{j_{4}}\,\ket{\Phi}\,
\bra{s}\,\anni{a}{j_{2}}\,
\anni{a}{j_{3}}\ket{\Phi}
\label{B16}
\end{eqnarray}
 with $\bra{s}\,\anni{a}{j_{1}}\,
\anni{a}{j_{2}}\,\ket{\Phi}\:=\:
\tilde{F}^{-1}_{j_{2}j_{1}}
\theta(j_{1})\,\theta(j_{2})$
 which also follows from (\ref{B15}).

\pagebreak

\noindent{{\bf  Figure Captions}

{\bf Figure 1.}
 Local average density profile at large times. 
 In the horizontal axis
 we display $y$ and in the vertical axis
 the density is given in units
 of $\frac{1}{2\sqrt{\pi t}}$.}
\begin{figure}[p]
\epsfbox{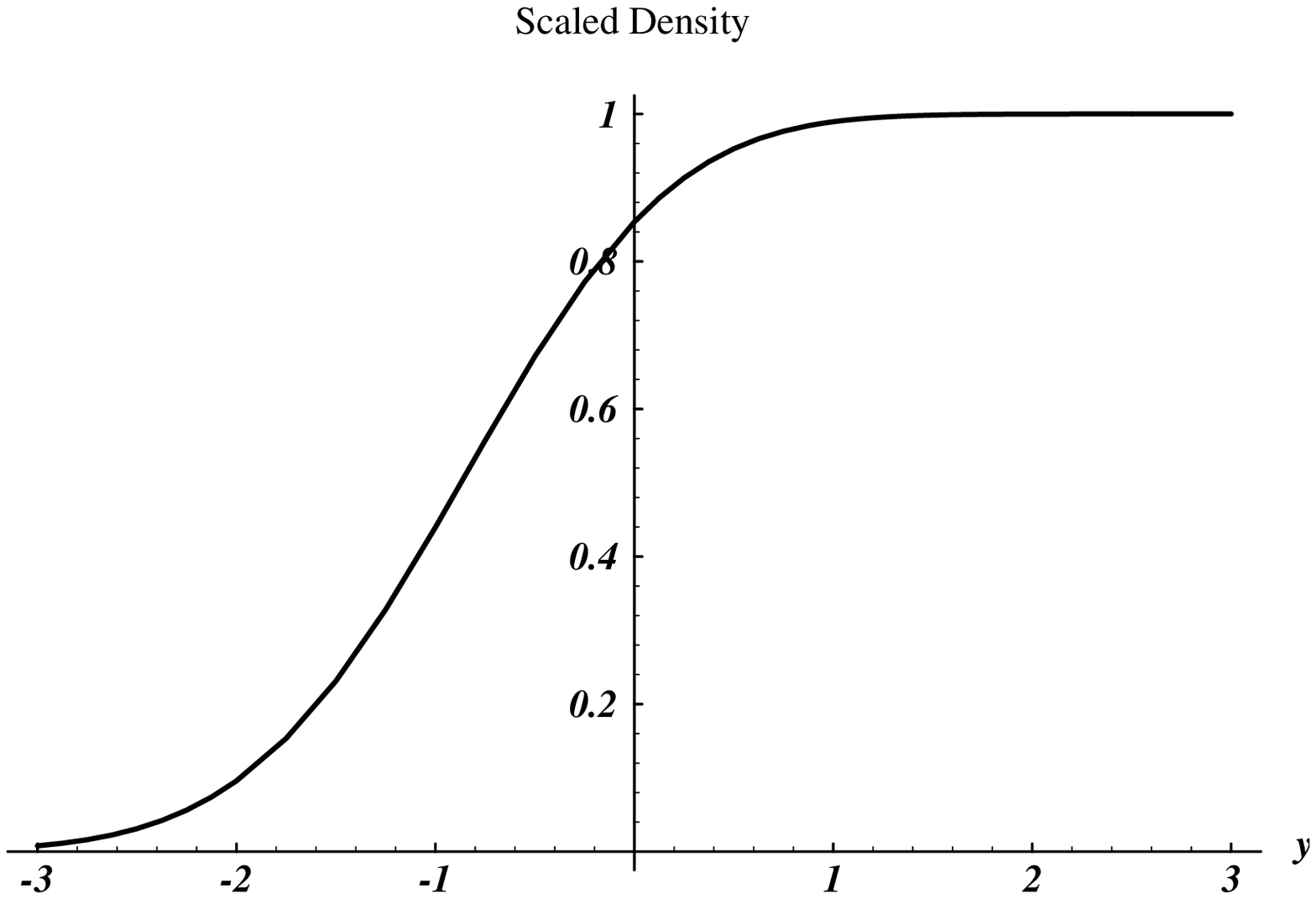}
\end{figure}

\end{document}